\documentclass[conference]{IEEEtran}

\IEEEoverridecommandlockouts
\usepackage{algorithm}
\usepackage{amsmath}
\usepackage{algpseudocode}
\usepackage{cite}
\usepackage{amsmath,amssymb,amsfonts}
\usepackage{graphicx}
\usepackage{subcaption}
\usepackage{textcomp}
\usepackage{xcolor}
\usepackage{amsmath,amssymb,amsfonts}
\usepackage{graphicx}
\usepackage{textcomp}
\usepackage{xcolor}
\usepackage{cases}

\usepackage{cancel}

\def\BibTeX{{\rm B\kern-.05em{\sc i\kern-.025em b}\kern-.08em
    T\kern-.1667em\lower.7ex\hbox{E}\kern-.125emX}}
\begin{document}

\title{
Towards Cloud-Native Agentic Protocol Learning for Conflict-Free 6G: A Case Study on Inter-Slice Resource Allocation
\\
\author{Juan Sebastián Camargo$^{(1)}$, Farhad Rezazadeh$^{(2)}$, Hatim Chergui$^{(1)}$, Shuaib Siddiqui$^{(1)}$ and Lingjia Liu$^{(3)}$\\
{\normalsize{} $^{(1)}$ i2CAT Foundation, Barcelona, Spain}\\{\normalsize{} $^{(2)}$ Centre Tecnol\'ogic de Telecomunicacions de Catalunya (CTTC), Barcelona, Spain}\\{\normalsize{} $^{(3)}$ Virginia Tech, Blacksburg, USA}\\
{\normalsize{}Contact Email: \texttt{\{name.surname\}@i2cat.net},~\texttt{frezazadeh@cttc.es},~\texttt{ljliu@vt.edu}}}
}

\IEEEoverridecommandlockouts

\definecolor{darkbrown}{rgb}{0.6, 0.26, 0.13}
\addtolength{\topmargin}{0.05in}

\clubpenalty=10000
\widowpenalty=10000

\maketitle

\begin{abstract}
In this paper, we propose a novel cloud-native architecture for collaborative agentic network slicing. Our approach addresses the challenge of managing shared infrastructure, particularly CPU resources, across multiple network slices with heterogeneous requirements. Each network slice is controlled by a dedicated agent operating within a Dockerized environment, ensuring isolation and scalability. The agents dynamically adjust CPU allocations based on real-time traffic demands, optimizing the performance of the overall system. 
A key innovation of this work is the development of emergent communication among the agents. Through their interactions, the agents autonomously establish a communication protocol that enables them to coordinate more effectively, optimizing resource allocations in response to dynamic traffic demands. Based on synthetic traffic modeled on real-world conditions, accounting for varying load patterns, tests demonstrated the effectiveness of the proposed architecture in handling diverse traffic types, including eMBB, URLLC, and mMTC, by adjusting resource allocations to meet the strict requirements of each slice. Additionally, the cloud-native design enables real-time monitoring and analysis through Prometheus and Grafana, ensuring the system's adaptability and efficiency in dynamic network environments. The agents managed to learn how to maximize the shared infrastructure with a conflict rate of less than 3\%. 

\end{abstract}

\begin{IEEEkeywords}
6G, Multi-agent Reinforcement Learning, Emerging Communication, Network Slicing, Shared Infrastructure
\end{IEEEkeywords}

\section{Introduction}

Recent studies \cite{9432398, Miuccio2022LearningGW, miuccio2023learning, Marwa_Chafii_2023} have made significant progress in communication protocols, negotiation, adaptability, and emergent communication for next-generation networks. For instance, \cite{9432398} demonstrated how deep reinforcement learning (DRL) agents can autonomously develop medium access control (MAC) protocols. Building on this, \cite{Miuccio2022LearningGW} and \cite{miuccio2023learning} explored advanced architectures incorporating autoencoders and long short-term memory (LSTM) networks, achieving enhanced robustness and faster convergence in 6G protocol learning. Furthermore, \cite{Marwa_Chafii_2023} highlighted the importance of emergent communication in dynamic 6G systems, while \cite{intelligible} introduced a protocol learning framework to minimize inter-slice conflicts and underutilization in resource allocation.

As networking environments grow more complex, with diverse tasks competing for limited resources, Multi-Agent Deep Reinforcement Learning (MADRL) \cite{albrecht2024multi} emerges as a promising solution. By enabling multiple agents to collaboratively optimize shared resources, MADRL has demonstrated potential in energy-efficient cloud systems \cite{WANG2025103671, 10417832} and Radio Access Network (RAN) slicing \cite{ranslicing, 10539464}. Each MADRL agent independently learns to manage its slice's resources while dynamically adapting to fluctuating traffic and coordinating with other agents. This capability is especially critical in B5G/6G networks, where slicing creates isolated, virtualized network segments tailored to specific services.

A unique aspect of collaborative agentic systems is emergent communication, where agents autonomously develop protocols to exchange vital information without revealing sensitive details. This ability enhances coordination for tasks like spectrum sharing and load balancing, contributing to more efficient and resilient network operations \cite{10488259}. Our work builds on these advances by integrating MADRL in an edge cloud setup, where each agent represents a slice-specific intelligent module utilizing inter-agent emergent communication for resource optimization.

In this paper, we propose a novel approach to maximize shared infrastructure by reallocating excess capacity from underutilized slices to those in need. Agents autonomously manage their slice's CPU allocation and coordinate using learned numerical communication protocols. These emergent protocols, developed during training, enable agents to optimize resource sharing effectively while maintaining slice isolation.

Our contributions include i) a cloud-native Dockerized MADRL architecture for dynamic resource allocation in B5G/6G slices; ii) integration of emergent communication for decentralized coordination in an agentic system via a Kafka bus; and iii) a demonstration of performance gains, including reduced conflicts, higher resource utilization, and improved latency management.
This approach not only addresses current challenges in network slicing but also lays the foundation for scalable, autonomous optimization in future network architectures.

\section{Environment Description}
\subsection{System Model}

We consider a 6G RAN-Edge architecture that simultaneously serve a set $\mathcal{K}$ of $K$ network slices ($k=1\ldots,K$) with heterogeneous requirements. The Edge domain is characterized by its total CPU frequency $f_{\max}$ and number of processed bits per CPU cycle $U$ and, for each slice $k$, runs an instance of application server (APP) as well as a user-plane function (UPF) as CNFs. Both RAN and Edge technological domains are controlled by separate agents per slice, which are allowed to communicate with each other via control messages, without exchanging monitoring data.

\begin{figure*}[t]
    \centering
    \includegraphics[width=0.8\linewidth]{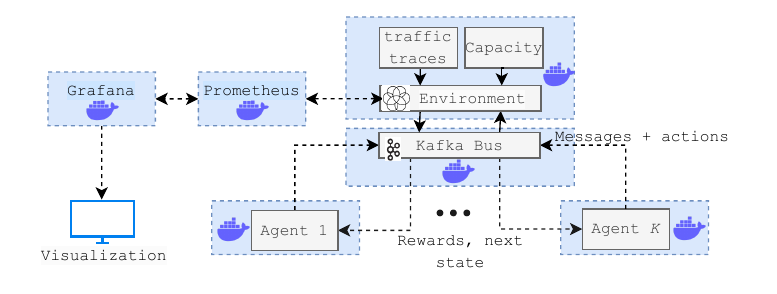}
    \caption{Docker-based protocol learning architecture.}
    \label{fig:docker_arch}
\end{figure*}
At time interval $t$ ($t=1\ldots,T$) of granularity $\tau$, downlink service packets are requested to slice $k$ application server located at the Edge according to a dynamic Poisson traffic arrival rate,
\begin{equation}
\lambda_k^{(t)} = \max\{\lambda \sim \mathcal{N}(\mu_k, \sigma_k), 0\}.     \label{lambda}
\end{equation}
In practice, the distribution parameters are determined based on real traffic samples.
\subsection{Edge-RAN Queuing Model}

The UPF virtual instance $k$ is endowed with a computation queue to locally process the packets arriving from slice $k$ Edge application server before sending them to the RAN CU which implements a communication queue to transmit packets through the radio interface. 

Specifically, upon packet requests at time $t$, $U_{t}^{(a)}$ server bits are delivered to UPF $k$ according to the dynamic Poisson traffic profile (\ref{lambda}). The computation queue is therefore drained as,
\begin{equation}
    Q_{k,t+1}^{(e)} = \max{\left(0, Q_{k,t}^{(e)} -  U_{k,t}^{(e)}\right)}+ U_{k,t}^{(a)}, \quad k \in \mathcal{K},
\end{equation}
wherein $U_{k,t}^{(e)}=\tau f_{k,t} U$ denotes the number of processed bits during interval $t$, which depends on the CPU frequency $f_{k,t}$ (in CPU cycles per second) allocated to slice $\ k \in \mathcal{K}_k$. Next, the RAN CU communication queue is drained as,
\begin{equation}
\begin{split}
\label{Eqn:computation_latency}
Q_{k,t+1}^{(r)} &= \max{\left(0, Q_{k,t}^{(r)} - U_{k,t}^{(r)}\right)} \\
     &+ \min{\left(Q_{k,t}^{(r)}, U_{k,t}^{(e)}\right)},\, k \in \mathcal{K},
\end{split}
\end{equation}
where $U_{k,t}^{(r)}$ denotes the number of transmitted bits over the radio interface during interval $t$, which depends on the channel capacity $C_{k,t}$, i.e.,
 \begin{equation}
     U_{k,t}^{(r)} = \Delta t \times C_{k,t}.
 \end{equation}

Therefore, the aggregation of the Edge UPF computation and RAN CU communication queues yield,
\begin{equation}
 Q_{k,t} = Q_{k,t}^{(e)} + Q_{k,t}^{(r)} \quad k \in \mathcal{K}, \label{tot_queue}
\end{equation}
where we omit both the backhaul and fronthaul delays since they are negligible constants in the considered scenario.

\subsection{Slice-Level Latency}
According to Little's law \cite{little}, the latency is proportional to the average queue length (\ref{tot_queue}). Specifically, by considering the average traffic arrival rate $\bar{U}_{k}^{(a)} = \mathbf{E}_{t'}\left[U_{k,t'}^{(a)}/\tau\right]$, user $k$ end-to-end long-term latency until time step $t$ can be expressed as,
\begin{equation}
    L_{k,t} = L_{k,t}^{(e)} + L_{k,t}^{(r)}, \label{E2ELatency}
\end{equation}
where the Edge and RAN latencies can be written respectively as,

\begin{subequations}
    \begin{equation}
         L_{k,t}^{(e)} = \bar{U}_{k}^{(a)} \times \frac{1}{t}\sum_{t'=1}^{t}Q_{k,t'}^{(c)},
    \end{equation}
    \begin{equation}
        L_{k,t}^{(r)} = \bar{U}_{k}^{(a)} \times \frac{1}{t}\sum_{t'=1}^{t}Q_{k,t'}^{(r)}.
    \end{equation}
\end{subequations}
Indeed, Little's Law is used to translate the complex dynamics of task processing and queuing into a simple yet powerful metric (latency) that the reinforcement learning agents can use to improve their decision-making processes.

On the other hand, slice $k$ Edge agent is aiming at optimizing the end-to-end SLA by allocating CPU resources in such a way to improve the computing queue, while ensuring that the available CPU capacity at the Edge $f_{\max}$ is not exceeded; the event where slice $k$ would be penalized and allocated its isolation CPU share $f_{th,k}$ if it was exceeding it.

\subsection{Utilization}

Regarding the utilization in a model-free scenario, slices operate independently without coordination or communication as the resource allocation is based on a static policy. Therefore, having a traffic $\Phi_{k,t}$ for any given slice $k$ at time $t$, the total utilization $Z_t$ is given by,

\begin{equation}
         Z_t = \frac{\sum_{k=1}^{K} \Phi_{k,t}}{U \times f_{max}}.
\end{equation}   

\section{Cloud Native Architecture}

Cloud-native designs align well with the evolving needs of 5G and future 6G networks, which demand high flexibility, scalability, and resilience. By leveraging containerization, microservices, and continuous monitoring, the proposed architecture in Fig.~\ref{fig:docker_arch} can efficiently support the diverse and stringent requirements of different network slices, ranging from ultra-reliable low-latency communications (URLLC) to enhanced mobile broadband (eMBB) and massive machine-type communications (mMTC). This modular and scalable architecture is also future-proof, enabling seamless integration with emerging technologies and facilitating the transition towards fully autonomous and self-optimizing networks in the 6G era. The different blocks of the architecture are detailed in the sequel.

\subsection{Agent Containers}
Indeed, we propose a cloud-native architecture to manage network slices in future 6G environments, utilizing a collaborative multi-agent communication framework. Each network slice is controlled by an independent agent that operates within a containerized environment, using Docker as the application abstraction layer. The containerization of these agents ensures that each operates in an isolated environment, providing a lightweight, scalable, and efficient mechanism to manage resources across multiple slices. Dockerization also facilitates the deployment, scaling, and orchestration of these agents across a distributed cloud infrastructure, making it easier to adapt to varying workloads and network demands inherent in 5G and 6G use cases.
\subsection{Server Container}
The agents communicate with a central server that functions as the information exchange hub. This centralized server ensures that while agents collaborate to optimize the infrastructure's overall performance, they do so without directly sharing sensitive information with each other. This design is crucial in maintaining privacy and security, as each agent remains unaware of the specific resource allocations and operational strategies of other agents. The central server aggregates and processes the data, ensuring that the collaborative decision-making process respects the constraints of the shared infrastructure and adheres to the privacy requirements of each network slice.
\subsection{Communication through Kafka Bus}
Agent-to-agent and agent-to-server communication occurs via a dedicated Kafka Bus, a distributed streaming platform designed for real-time, high-throughput, and low-latency data exchange. Kafka serves as the backbone of the MARL environment, enabling asynchronous communication crucial for emergent protocols and dynamic interactions.

The centralized server uses Kafka to receive state updates and action data from agents, calculate rewards, and coordinate network configurations. This decoupled architecture ensures agents and the server operate independently, maintaining system efficiency and scalability. Kafka's robustness, fault tolerance, and scalability make it an ideal choice for simulating complex, distributed environments and supporting the evolving communication strategies of MARL agents.
\begin{figure*}[t]
    \centering
    \begin{subfigure}[h]{0.48\textwidth}  
        \centering
        \includegraphics[width=\textwidth]{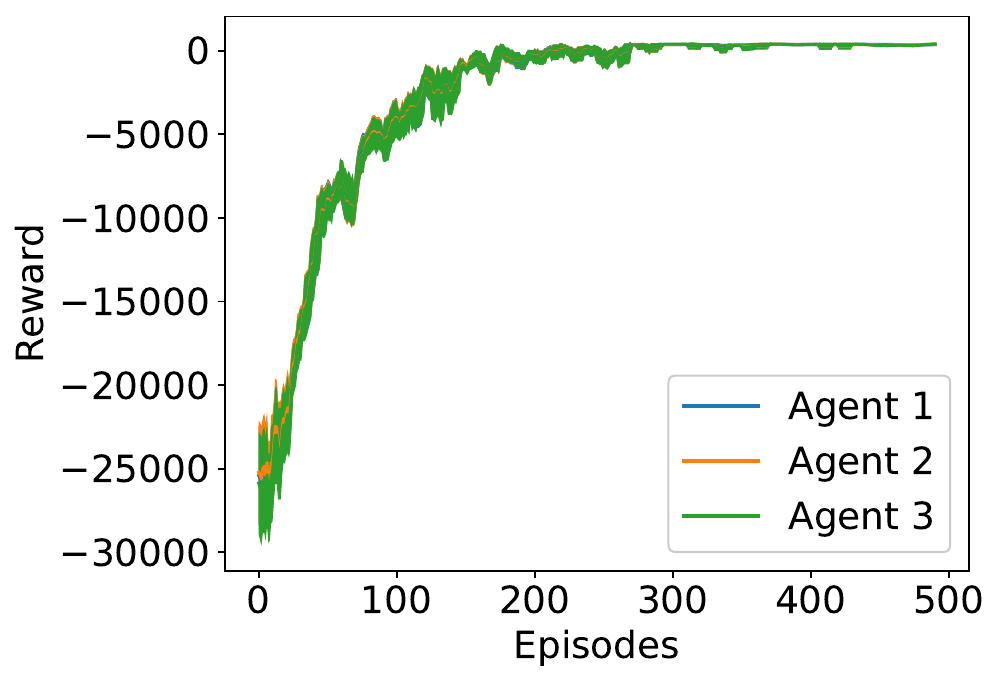}
        \caption{Average reward throughout training.}
        \label{fig:plotA}
    \end{subfigure}
    \hspace{0.1cm}  
    \begin{subfigure}[h]{0.45\textwidth}  
        \centering
        \includegraphics[width=\textwidth]{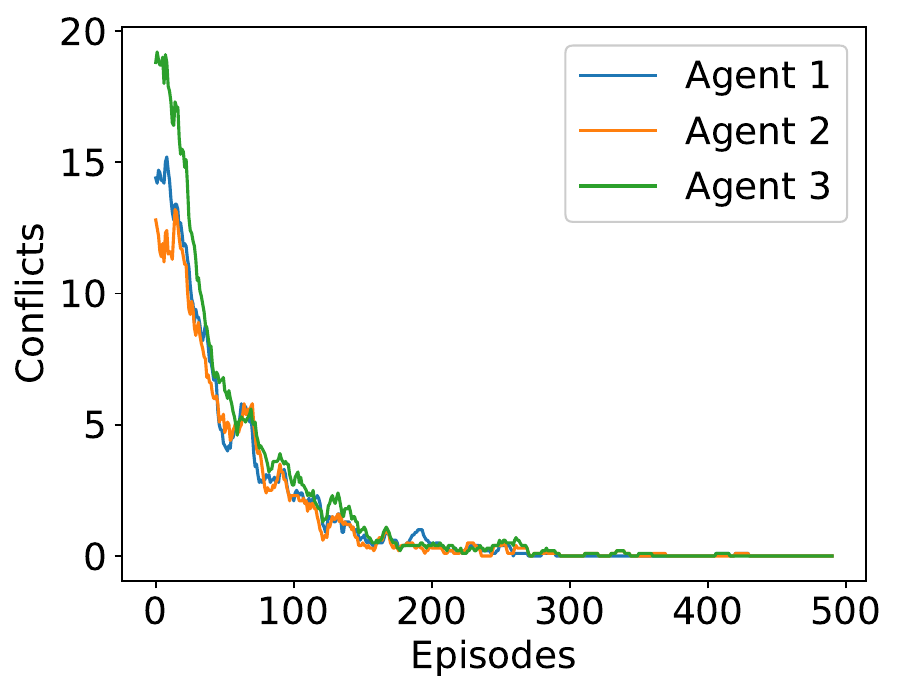}
        \caption{Average conflict throughout training.}
        \label{fig:plotB}
    \end{subfigure}
    
    \caption{Average reward and average conflict throughout the training process.}
    \label{fig:combined_plots}
\end{figure*}

\subsection{Monitoring via Prometheus and Grafana}

A key advantage of this cloud-native architecture is its ability to be monitored and managed in real-time using advanced monitoring and visualization tools like Prometheus and Grafana. Prometheus collects detailed metrics from each agent, allowing for real-time performance analysis, anomaly detection, and resource optimization. Grafana, in turn, provides a user-friendly interface for visualizing these metrics, offering insights into the operational status of the entire system. This online supervision capability is particularly valuable in dynamic B5G and 6G environments, where network conditions and demands can change rapidly, requiring swift and informed decision-making.

\section{MA-DRL Communication Formulation}

In an edge environment with shared infrastructure, we have $K$ slices (agents) that need to allocate resources efficiently. Each slice represents a type of application or service running on the shared infrastructure. The goal is to optimize resource allocation among these slices while adhering to certain constraints, ensuring that all slices can operate effectively, maximizing the infrastructure resources but without exceeding the total computing capacity. 

We define  $a_{k,t}$ as the amount of resources allocated to slice $k$ at time $t$. The emergent communication protocol, each slice $k$ learns a policy $\pi_k$ that represents its selected message and actions based on the observations and the received messages. The emergent communication protocol evolves during training as agents interact and share information to coordinate the resources. This policy is affected by the message $m_{k,t}$, which as stated before, is a discrete value. Therefore, at any given time $t$, the slice $k$ sends a message $m_{k,t} \in \mathcal{M}_k$. As such, we can describe $\pi_k$ such as: 

\begin{equation}
         \pi_i(a_{k,t}, m_{k,t} \mid o_{k,t}, \bar{m}_{k,t}),
    \end{equation}

Where $\bar{m}_{k,t}$ represents the set of received messages from other agents and $o_{k,t}$ stands for the observations, which includes two metrics, namely, the normalized traffic $\bar{\Phi}_{k,t}$ and the CPU allocation gap, i.e., $\gamma_k = f_{th,k} - a_{k,t}$.

The objective of the policy is to maximize the efficiency of the resource utilization and penalize conflicts. In this regard, the utilization function $Z_{k,t}$ for each slice is defined as the ratio between the used resources (as a function of the incoming traffic) and the allocated resources by the agent of each slice $k$, i.e.,

\begin{equation}
\label{util}
    Z_{k,t} = \frac{\Phi_{k,t}}{U \times a_{k,t}}.
\end{equation}

At the same time, we need to restrict the sum of allocated CPU per slice, as they cannot exceed the total available CPU of the shared infrastructure, i.e., 

\begin{equation}
         \sum_{k=1}^{K} a_{k,t} \leq f_{max}.
    \end{equation}
In this respect, the situation where the concurrent allocations surpass the limit is termed \emph{conflict}. In this case the agents exceeding their default share $f_{th,k}$ are penalized through a negative reward $-\theta$. Otherwise, the reward is fostering the minimization of latency, i.e.,

\begin{equation}
 r_{k,t} =
\begin{cases}
     -\theta, & \text{if penalty}  \\
     \alpha \exp\{{-L_{k,t}}\}, & \text{else}   
\end{cases}
\end{equation}

Additionally, the resources should not be negative, so we need to restrict them also as: 

\begin{equation}
        a_{k,t} \geq 0, \quad \forall k \in \{1, \dots, K\}
    \end{equation}

Note that the reward function is linked with the agent's policy $\pi_i$ because the policy determines the actions and messages that influence the system state and the resulting rewards. By optimizing their policies, agents aim to maximize their expected cumulative rewards while coordinating with other agents through emergent communication. 

By incorporating the CPU gap and traffic levels into the agent's observation space, and designing a reward function that penalizes conflicts while prioritizing latency minimization, the agent naturally aligns its actions with traffic trends. This approach would enable the agent to allocate only the necessary CPU resources, optimizing their usage through the established link in eq. (\ref{util}). As a result, the multi-slice system achieves higher resource utilization and experiences reduced conflicts, fostering a more efficient and harmonious operation.
\begin{figure*}[t]
    \centering
\includegraphics[width=0.7\linewidth]{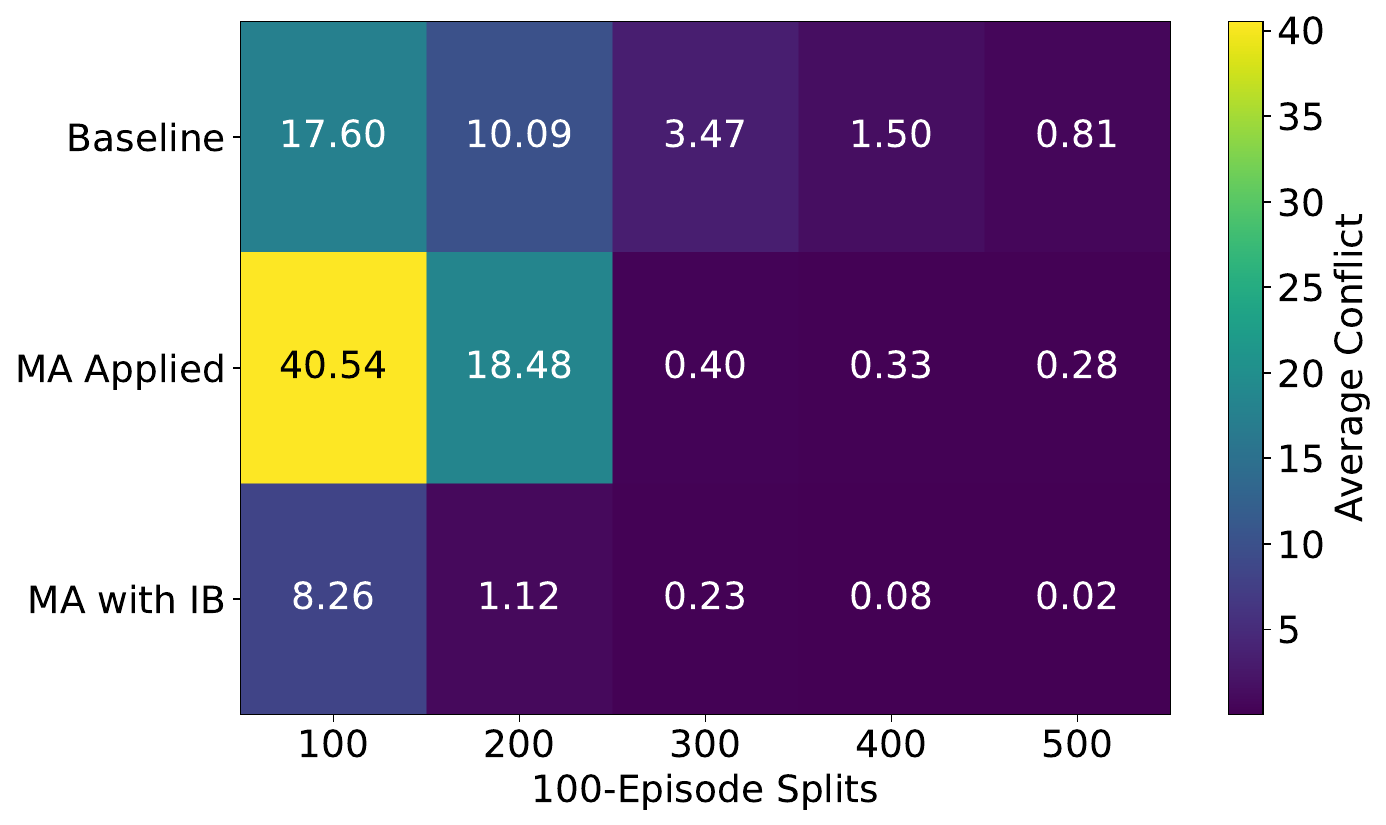}
    \caption{Conflicts heat map vs. different MA models.}
    \label{fig:MA_comparison}
\end{figure*}
\section{Results}

We tested our model with a synthetic traffic and using the Cloud architecture described in Section III. For illustration purposes, we created an environment with three agents. We then ran the environment for 500 episodes and analyzed the results. For the performance evaluation, the system's traffic was modeled based on real-world network conditions, captured in the traffic dataset used during the simulation, as described in \cite{rezazadeh2022specialization}. It has a granularity of $\tau=10$ ms. This approach enabled the simulation to account for varying traffic loads and conditions across different times of the day and different locations within the city, providing a comprehensive testing environment for evaluating the performance of the proposed multi-agent control model. The maximum CPU frequency for the edge server hosting the slices was set at 40 Giga cycle/s (multi-core), a limit that represents the total computational capacity available to the agents. The default CPU shares in Giga cycle/s for the three slices are $f_{th} = [15, 15, 10]$, $U=10^{-10}$ bits per CPU cycle. This parameter determined the rate at which each agent could handle incoming traffic, directly influencing queue lengths, latency, and overall system performance. Three distinct types of traffic were reflected in the packet size parameter: eMBB with 1500-bit packets, URLLC with 32-bit packets, and mMTC with 40-bit packets. These variations represent the diverse requirements of different network services, from high-throughput applications to low-latency, high-reliability scenarios. The different packet sizes resulted in varying computational demands, with larger packets requiring more processing power, thereby influencing the agents' actions and the overall system dynamics. All the messages belong to the finite set $\mathcal{M}=\{1,2,3\}$.
\subsection{Monitored Metrics via Grafana}
The messages are exchanged using the Kafka Server bus, and we are monitoring the performance of the AI module using a combination of Prometheus and Grafana servers. The metrics that we monitor includes the rewards of each cycle and the number of conflicts among the slices, as can be seen in Figure \ref{fig:plotA} and \ref{fig:plotB}, respectively.

\subsection{Conflicts vs. episodes}
As shown in Figure \ref{fig:MA_comparison}, we compare three different versions of the MA model under identical traffic and configuration settings, with the only variation being the architecture of the MA-DQN. The first version, serving as the baseline, is a vanilla MA-DQN configuration without emergent communication. The second version (MA Applied) incorporates emergent communication but lacks the prioritized replay buffer and information bottleneck (IB). The third version (MA with IB) includes all features: emergent communication, the prioritized replay buffer, and the IB.

The results highlight the significant performance advantages of MA with IB, which achieves near-zero conflicts on average by the end of the training cycle. The impact of emergent communication is particularly evident, substantially reducing conflicts compared to the baseline. While the models with emergent communication exhibit a slower start during the early training stages, this initial delay is quickly overcome. Midway through training, the models using emergent communication outperform the baseline by a considerable margin, demonstrating the critical role of this feature in improving overall performance.

\subsection{Latency}

\begin{figure}
    \centering
\includegraphics[width=1.0\linewidth]{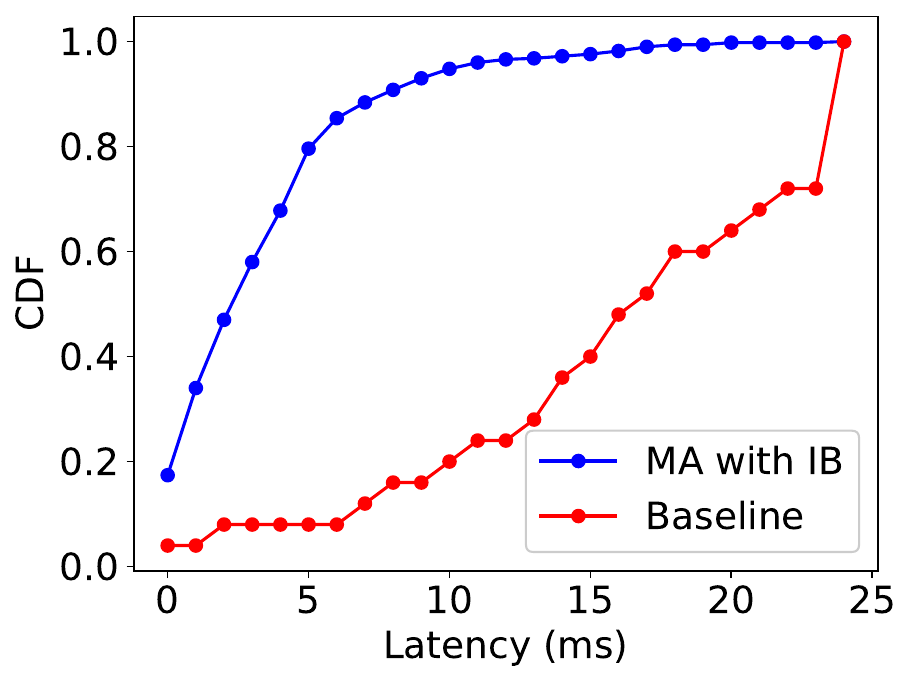}
    \caption{Latency cumulative distribution function for MA IB}
    \label{fig:deep_q_learning_cycle}
\end{figure}

Additionally, Figure \ref{fig:deep_q_learning_cycle} represents the Cumulative Distribution Function (CDF) of the observed latency (in milliseconds) for a specific agent during training. The CDF shows the proportion of episodes where the latency remains below a given threshold, offering an insightful view of the agent's performance in terms of delay-sensitive tasks.

We observe that the CDF starts at approximately 0.4 and quickly rises, reaching a value close to 1.0. This indicates that around 90\% of the latency observations are below the 200 ms threshold. The steep initial slope signifies that the majority of latency values are clustered around lower latency levels, which is a desirable property for latency-sensitive applications. The CDF highlights the effectiveness of the reinforcement learning model in managing latency within acceptable thresholds for a significant proportion of episodes. The steep increase in the CDF curve reflects that the model is able to maintain low-latency performance for the majority of the agent's interactions, demonstrating its robustness in real-time decision-making scenarios.


\subsection{Utilization}
\begin{figure}[t]
    \centering
\includegraphics[width=1.0\linewidth]{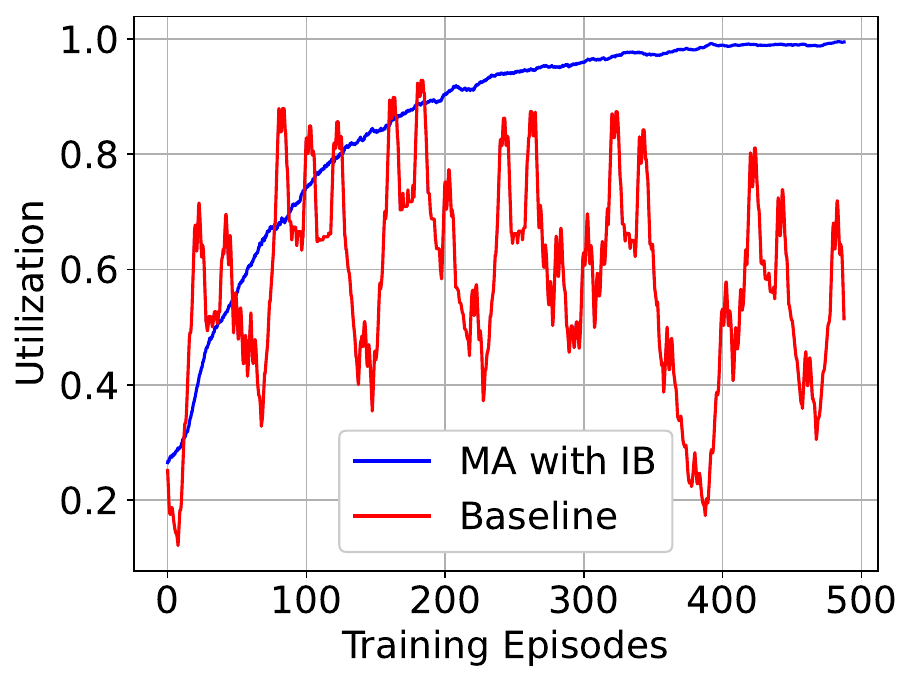}
    \caption{Utilization vs. episodes}
    \label{fig:utilization_graph}
\end{figure}

Another key factor to consider is the utilization of the shared infrastructure. We define utilization as the ratio between the maximum available processing power and the current capacity in use. By monitoring the infrastructure's utilization, we can assess how effectively the MARL agents optimize resource allocation compared to a system configuration without the model. The utilization through training episodes can be seen in Figure \ref{fig:utilization_graph}. As the training progresses, the models can distribute the computational process more efficiently between the slices, obtaining a better use of the infrastructure compared with a fixed approach. We have reached almost an 85\% of infrastructure utilization while using the multi-agent slicing model, compared with an average of just 57\% of baseline approach. This improvement in the utilization brings several positive points in the network, such as a decrease in over-provisioning, obtaining a better value out of the investment of infrastructure and improving the energy efficiency of the systems overall.

\section{Conclusion}
In this paper, we have introduced a novel inter-slice conflict and underutilization minimization framework based on the concept of Multi-Agent protocol learning. By letting each slice agent exploit information bottleneck auto-encoding to extract the latent variables in the inter-agent emergent communication messages, the conflicts have been dramatically reduced and the utilization of the shared infrastructure has been notably improved via adaptation to fluctuating traffic and network conditions. The framework has been implemented in a cloud-native environment using Dockerized containers and synthetic traffic designed that mimics real-world scenarios. After the training phase, the agent system has successfully developed a new communication protocol, optimizing the use of shared resources while minimizing conflicts between them.
\vspace{5mm}
\section{Acknowledgements}
This paper has been partially supported by MCIN/AEI/ 10.13039/501100011033 (ERDF “a way of making Europe”) under grant PID2022-142332OA-I00 and the Horizon Europe projects' NANCY (101096456) and COGNIFOG (101092968).
\bibliographystyle{IEEEtran}
\bibliography{biblio.bib}
\end{document}